\documentclass[pra,aps,showpacs,preprint]{revtex4}
\usepackage{amsmath}
\usepackage{graphicx}
\usepackage{dcolumn}
\usepackage{bm}

\begin{document}
\title{Mean-field dynamics of a Bose Josephson junction in an optical cavity}
\author{J.~M. Zhang, W.~M. Liu, and D.~L. Zhou}
\affiliation{Beijing National Laboratory for Condensed Matter
Physics, Institute of Physics, Chinese Academy of Sciences, Beijing
100080, China}
\begin{abstract}
We study the mean-field dynamics of a Bose Josephson junction which
is dispersively coupled to a single mode of a high-finesse optical
cavity. An effective classical Hamiltonian for the Bose Josephson
junction is derived and its dynamics is studied in the perspective
of phase portrait. It is shown that the strong condensate-field
coupling does alter the dynamics of the Bose Josephson junction
drastically. The possibility of coherent manipulating and \textsl{in
situ} observation of the dynamics of the Bose Josephson junction is
discussed.
\end{abstract}
 \maketitle Cavity
quantum electrodynamics has now grown into a paradigm in the study
of matter-field interaction. To tailor the atom-field coupling
effectively, high degree of control over the center-of-mass motion
of the atoms are essential. Although previous works have focused on
the few-atom level \cite{kimble,rempe,chapman2}, recently, a great
step was made as two groups succeeded independently in coupling a
Bose-Einstein condensate to a single cavity mode \cite{esslinger,
colombe}. That is, a single cavity mode dressed condensate has been
achieved. This opens up a new regime in both the fields of cavity
quantum electrodynamics and ultracold atom physics. In the
condensate, all the atoms occupy the same motional mode, and couple
identically to the cavity mode, thus realizing the Dicke model
\cite{dicke} in a broad sense. As shown by these experiments, the
condensate is quite robust, it would not be destroyed by its
interaction with the cavity mode.

In this paper, we investigate the mean-field dynamics of a Bose
Josephson junction (BJJ), which is coupled to a driven cavity mode.
This extends our previous work to the many-atom case \cite{zhang}.
The system may be constructed by splitting a Bose Einstein
condensate, which already couples to a single cavity mode, into two
weakly linked condensates, as can be done in various ways
\cite{hofferberth,levy, schumm, albiez, shin}. We restrict to the
large detuning and low excitation limit, so that the atomic
spontaneous emission can be neglected. In this limit, the effect of
the strong coupling between the atoms and the field, seen by the
field, is to shift the cavity resonance frequency and hence modifies
the field intensity. Unlike the single condensate case, we now have
two, which may couple with different strengths to the cavity mode
because of the position dependence of the atom-field coupling.
Consequently, the field dynamics is coupled to the tunneling
dynamics of the BJJ, and vice versa. It is the interplay between the
two sides that we are interested in. We would like to stress that,
although there had been some experimental investigations on this
subject and phenomena such as dispersive optical bistability were
observed \cite{hemmerich,gupta}, all of them dealt with thermal cold
atoms. However, here, the long coherence time of the condensates
will surely make a difference.

The Hamiltonian of the system consists of three parts:
\begin{equation}\label{H}
    H=H_a+H_f+H_{\mathrm{int}}.
\end{equation}
$H_a$ is the canonical Bose Josephson junction Hamiltonian in the
two-mode approximation ($\hbar=1$ throughout),
\begin{equation}\label{ha}
    H_a=-\Omega (b_1^\dagger b_2+b_2^\dagger b_1
    )+\frac{V}{2}(b_1^\dagger b_1^\dagger b_1 b_1 +b_2^\dagger b_2^\dagger b_2
    b_2),
\end{equation}
where $b_1^\dagger$, $b_2^\dagger$ ($b_1$, $b_2$) create
(annihilate) an atom in its internal ground state in the left and
right trap respectively. $\Omega$ is the tunneling matrix element
between the two modes, while $V$ denotes the repulsive interaction
strength between a pair of atoms in the same mode. The two-mode
model assumes two stationary wave functions (the single-atom ground
states actually) in the individual traps, while neglecting the
modifications due to the atom-atom interaction. The regime in which
this is the case can be found in Ref.~\cite{walls}. $H_f$ is the
single-mode field Hamiltonian,
\begin{equation}\label{hf}
    H_f=\omega_c a^\dagger a+\eta (t)e^{-i\omega_p t}a^\dagger+ \eta^\ast (t)e^{i\omega_p
    t}a,
\end{equation}
where $\omega_c$, $\omega_p$ are the cavity mode frequency and pump
frequency respectively, and $\eta(t)$, the amplitude of the pump,
varies slowly in the sense that $|\dot{\eta}/\eta|\ll \omega_p$. In
the limit of large detuning \cite{detuning} and weak pump, the atom
field interaction is of dispersive nature, and the two-level atoms
can be treated as scalar particles with the upper level being
adiabatically eliminated. Under the two-mode approximation for the
atoms, the interaction between the atoms and the cavity mode is
\cite{ritsch1},
\begin{equation}\label{Hint}
    H_{\mathrm{int}}=U_0 a^\dagger a (J_1 n_1+J_2n_2),
\end{equation}
where $U_0=g_0^2/(\omega_c-\omega_a)$ is the light shift per photon,
i.e. the potential per photon an atom feels at an antinode, $g_0$
being the atom-field coupling strength at an antinode. The two
dimensionless parameters $J_{1,2}$ ($0\leq J_{1,2}\leq 1$) measure
the overlaps between the atomic modes with the cavity mode
\cite{zhang}. $n_1=b_1^\dagger b_1$ ($n_2=b_2^\dagger b_2$) counts
the atoms in the left (right) trap. The term $H_{\mathrm{int}}$ has
a simple interpretation, from the point of view of the cavity mode,
its frequency is renormalized; while from the point of view of the
atom ensemble, the trapping potential is tilted provided $J_1\neq
J_2$.

According to the Heisenberg's equation, we have
\begin{subequations}
\begin{eqnarray}
   i\dot{b_1} &=& -\Omega b_2 +V b_1^\dagger b_1 b_1 +J_1U_0a^\dagger a b_1,  \label{b1}\\
  i\dot{b_2} &=& -\Omega b_1 +V b_2^\dagger b_2 b_2 +J_2U_0a^\dagger a b_2, \label{b2}\\
   i\dot{a} &=& \left[\omega_c+U_0(J_1n_1+J_2n_2)\right]a -i\kappa a +\eta(t)e^{-i\omega_p
  t}. \quad\label{amotion}
\end{eqnarray}
\end{subequations} Note that in Eq.~(\ref{amotion}) we have put in
the term $-i\kappa a$ to model the cavity loss, with $\kappa$ being
the cavity loss rate. Under the mean-field approximation, we treat
the operators $b_1$, $b_2$, and $a$ as classical quantities,
$b_1\sim\sqrt{N_1}e^{i\theta_1}$, $b_2\sim\sqrt{N_2}e^{i\theta_2}$,
$a\sim\alpha$. Here $N_1$, $N_2$ are respectively the numbers of
atoms in the left and right condensates, and $\theta_1$, $\theta_2$
are their phases. By taking the mean field approximation, we
actually confine to the so called Josephson regime as elaborated in
detail in Ref.~\cite{leggett}. This regime is defined as $1/N\ll
V/\Omega \ll N$, $N=N_1+N_2$ being the total atom number, and is
characterized by small quantum fluctuations both in the relative
phase $\phi\equiv\theta_2-\theta_1$ ($\Delta \phi \ll1$) and in the
populations $N_{1,2}$ ($\Delta N_{1,2}\ll \sqrt{N}$). This property
then justifies the mean field approximation. As shown in
Ref.~\cite{smerzi3}, the mean field predictions (self-trapping etc.)
are well recovered in a full quantum dynamics on a short time scale,
and their breakdown occurs only at a long time scale which increases
exponentially with the total atom number.

It is clear from Eq.~(\ref{amotion}) that the relaxation time scale
of the cavity mode is of order $1/\kappa$, which is much shorter
than the plasma oscillation period of a bare Bose Josephson junction
\cite{dressed} which, roughly speaking, is of order $1/\Omega$. In
fact, the typical values of $\kappa$ of high-finesse optical
cavities are of order $2\pi\times10^6$ Hz, while the experimentally
observed $\Omega$ is of order $2\pi\times10-2\pi\times10^2$ Hz
\cite{albiez}. This implies that the cavity field follows the motion
of the condensates adiabatically \cite{ritsch}, thus from
Eq.~(\ref{amotion}) we solve
\begin{equation}\label{asolve}
    \langle a\rangle=\alpha(t)=\frac{\eta(t)e^{-i\omega_p
    t}}{i\kappa+[\omega_p-\omega_c-U_0(J_1N_1+J_2N_2)]},
\end{equation}
and the photon number is
\begin{equation}\label{photonnumber}
     \langle a^\dagger a\rangle=|\alpha(t)|^2=\frac{|\eta(t)|^2}{\kappa^2+[\Delta-\delta
     U_0(N_1-N_2)/2]^2},
\end{equation}
where $\Delta\equiv\omega_p-\omega_c-(J_1+J_2)NU_0/2$, and
$\delta\equiv J_1-J_2$ is the coupling difference between the two
atomic modes to the cavity mode. We then see that with other
parameters fixed, the photon number depends only on the atom
population difference between the two traps. Moreover, the motion of
the cavity mode couples to that of the condensates only in the case
that the two traps are placed asymmetrically with respect to the
cavity mode such that $\delta$ is nonzero. Considering that the
cavity field intensity varies rapidly along the cavity axis (with
period $\lambda/2\sim 0.5$ $\mu$m) while rather smoothly in the
transverse plane (with mode waist $w\sim 10-25$ $\mu$m), and that
the extension of the condensates and the separation between them are
in between, it may be wise to create the coupling difference
$\delta$ by transverse rather than longitudinal position difference
between the two condensates. In the experiment of Colombe \textsl{et
al}. \cite{colombe}, the transverse position of the cigar-shaped
condensate, which is aligned parallel to the cavity axis, can be
adjusted in the full range of the cavity mode waist. On this basis,
the condensate may be split along its long axis, by using of the
radio-frequency-induced adiabatic potential \cite{hofferberth,
schumm}, which is also compatible with an atom chip, into two parts
offsetted in the transverse direction. For a mode waist $w=10$
$\mu$m, a separation $d=1$ $\mu$m is hopeful to create a coupling
difference $\delta=0.12$ if the two condensates are located near the
inflection point $x_c=w/2$ of the cavity field intensity.

By introducing the dimensionless parameter $z=(N_1-N_2)/N$, which
describes the population of the atoms between the two traps, we
rewrite the photon number (\ref{photonnumber}) as
\begin{equation}\label{f}
    |\alpha(z,t)|^2=\frac{A(t)^2}{(z-B)^2+C^2},
\end{equation}
where the three dimensionless parameters are defined as
$A(t)=\eta(t)/[\delta U_0 N/2]$, $B=\Delta/[\delta U_0 N/2]$, and
$C=\kappa/[\delta U_0 N/2]$. We may understand $A(t)$, $B$, and $C$
as the reduced pumping strength, reduced detuning, and reduced loss
rate, respectively. Equation (\ref{f}) implies that the photon
number, as a function of $z$, is a Lorentzian centered at $z_c=B$
and with width $2C$. Since $-1\leq z \leq 1$, to maximize the
influence of the condensates on the cavity field, it is desirable to
have $B$ within the same interval and $C\lesssim1$. Under these
conditions, the atomic motion is able to shift the cavity in or out
of resonance. Apart from the factor $\delta$, the latter condition
means that, the light shift per photon times the number of atoms
exceeds the resonance linewidth of the cavity, which has been
realized both with a ring cavity \cite{hemmerich} and with a
Fabry-Perot cavity \cite{gupta}.

In the following, we follow closely the line of Refs. \cite{smerzi,
smerzi2}. Substituting Eq.~(\ref{f}) into Eqs.~(\ref{b1}) and
(\ref{b2}), we find that the two equations can be rewritten in terms
of $z$ and the phase difference $\phi$ as
\begin{subequations}
\begin{eqnarray}
  \frac{dz }{dt}&=& - \sqrt{1-z^2}\sin\phi, \\
   \frac{d\phi}{dt} &=&
   \frac{z}{\sqrt{1-z^2}}\cos\phi+r z+\frac{\delta  U_0}{2\Omega}|\alpha(z,t)|^2,
\end{eqnarray}
\end{subequations}
where the time has been rescaled in units of the Rabi oscillation
time $1/(2\Omega)$, $2\Omega t\rightarrow t$. The dimensionless
parameter $r\equiv NV/(2\Omega)>0$ measures the interaction strength
against the tunneling strength. We further define a Hamiltonian
$H_c=H_c(z,\phi,t)$ in which $z$ and $\phi$ are two conjugate
variables, i.e., $\dot{z}=-\frac{\partial H_c}{\partial \phi}$,
$\dot{\phi}=\frac{\partial H_c}{\partial z}$. Such a Hamiltonian is
\begin{eqnarray}
    H_c(z,\phi,t)&=&-\sqrt{1-z^2} \cos\phi +\frac{1}{2}r
    z^2+\frac{\delta  U_0}{2\Omega}F(z,t),\quad\quad\label{Hc}\\
   F(z,t)&=&\frac{A(t)^2}{C}\arctan\left(\frac{z-B}{C}\right).\label{F}
\end{eqnarray}
The first two terms in Eq.~(\ref{Hc}) are the Hamiltonian of a bare
Bose Josephson junction as was first derived in Refs. \cite{smerzi,
smerzi2}. They describe the energy cost due to the phase twisting
between the two condensates and the atom-atom repulsion
respectively. The last term may be termed as a cavity field induced
tilt, as can be seen from its derivation. It reflects that the two
traps, which are originally symmetric, are now subjected to an
offset determined by the atom populations. In its nature, this term
is similar to the potential an atom feels when passing a cavity
adiabatically \cite{haroche}, with the variable $z$ playing the role
of the center-of-mass of the atom. The Hamiltonian can be made
explicitly time-dependent if the pump strength varies in time. This
may offer us a tool to coherently manipulate the motion of a Bose
Josephson junction \cite{chapman}. However, in this work, we
concentrate on the case that the pump strength is a constant,
$\eta(t)\equiv \eta$, so that the system is autonomous and the
Hamiltonian is conserved in time.

As a one degree-of-freedom Hamiltonian system and with the
Hamiltonian itself being a first integral, the system is integrable
and there is no chaos. The trajectory of the system in the phase
space (plane) follows the manifold (line) of constant energy. Thus
qualitatively speaking, the dynamics of the system is to a great
extent determined by the structure of its phase portrait, or more
specifically, the number of stationary points, their characters
(minimum, maximum, or saddle), and their locations. Before
proceeding forward, we have some remarks on the structure of the
phase space of the system and its implications. Superficially, the
Hamiltonian $H_c$ is defined on the rectangular domain, $-1\leq z
\leq 1$, $0 \leq \phi \leq 2\pi$. However, physically $\phi$ is
periodic in $2\pi$, and for $z=\pm1$, $\phi$ is not well defined, so
we should identify $(z,0)$ with $(z,2\pi)$ and collapse the lines
$(z=\pm1,\phi)$ to two points [Mathematically, this is justified by
the fact that $H_c(z,0)=H_c(z,2\pi)$, and $H_c(1,\phi)=C_1$,
$H_c(-1,\phi)=C_2$, with $C_1$, $C_2$ being two constants].
Therefore, the domain of the Hamiltonian or the phase space of the
system is homeomorphous to a sphere. The Euler's theorem for a
smooth function on a sphere states that the number of minima $m_0$,
the number of saddles $m_1$, and that of maxima $m_2$, satisfy the
relation $m_0-m_1+m_2=2$ \cite{euler}. This relation can be checked
in Fig.~\ref{fig1} below.

In the following, we explore the dynamics of a Bose Josephson
junction uncoupled or coupled to a cavity mode in the perspective of
phase portrait. This approach has the advantage that it captures the
whole information of the BJJ dynamics into one \cite{Buonsante}. As
a first step, we work out the stationary points of the system, which
are determined by the equations $\frac{\partial H_c}{\partial z}=0$,
$\frac{\partial H_c}{\partial \phi}=0$. The second equation implies
that $\phi=0$ or $\phi=\pi$. Substituting these two possible values
of $\phi$ into the first one, we have two equations of $z$
respectively,
\begin{subequations}
\begin{eqnarray}
   f_1(z)=r z+\frac{z}{\sqrt{1-z^2}}+\frac{\tilde{A}}{(z-B)^2+C^2}&=& 0, \quad\label{z1}\\
   f_2(z)= r z-\frac{z}{\sqrt{1-z^2}}+\frac{\tilde{A}}{(z-B)^2+C^2}&=&
  0. \quad\label{z2}
\end{eqnarray}
\end{subequations} where $\tilde{A}=\delta  U_0A^2/(2\Omega)$. The character (minimum, saddle, or maximum) of the possible stationary points
are determined by the corresponding Hessian matrices.

As a benchmark, we first consider the uncoupled case. If $r<1$,
there are a minimum $(z,\phi)=(0,0)$ and a maximum
$(z,\phi)=(0,\pi)$. If $r>1$, the point $(z,\phi)=(0,0)$ keeps to be
a minimum while $(z,\phi)=(0,\pi)$ turns now into a saddle point,
and there are two maximum at $(z,\phi)=(\pm\sqrt{r^2-1}/r,\pi)$. The
transition of the point $(z,\phi)=(0,\pi)$ from a maximum to a
saddle, and the split (bifurcation) of this old maximum into two new
maxima at $r=1$, mark the onset of running-phase and $\pi$-phase
self-trapping states \cite{smerzi2, walls}. Note that the
aforementioned Euler's theorem carries over from $r<1$ to $r>1$.
\begin{figure*}[thb]
\begin{minipage}[b]{0.45 \textwidth}
\includegraphics[width=\textwidth]{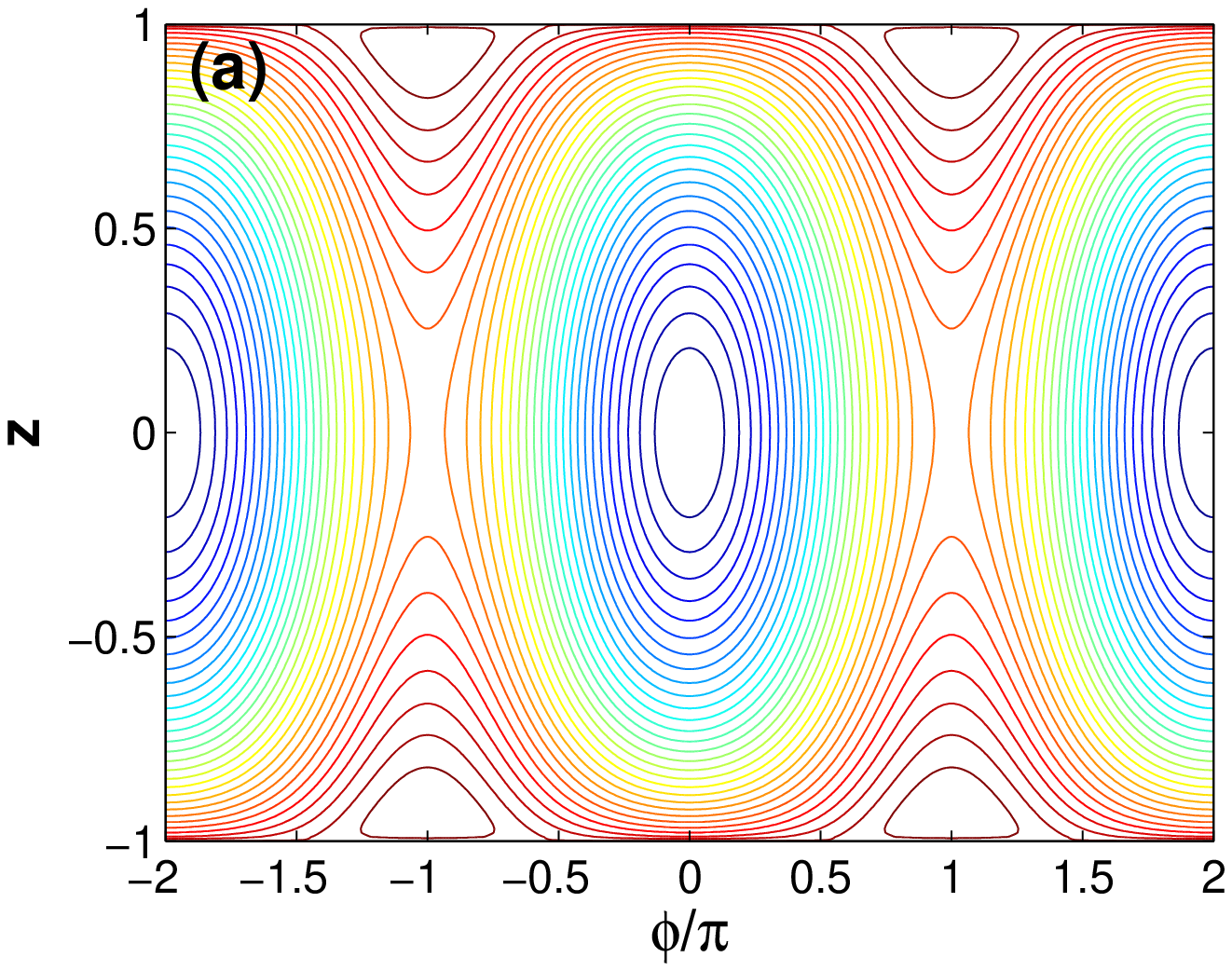}
\end{minipage}
\begin{minipage}[b]{0.45 \textwidth}
\includegraphics[width=\textwidth]{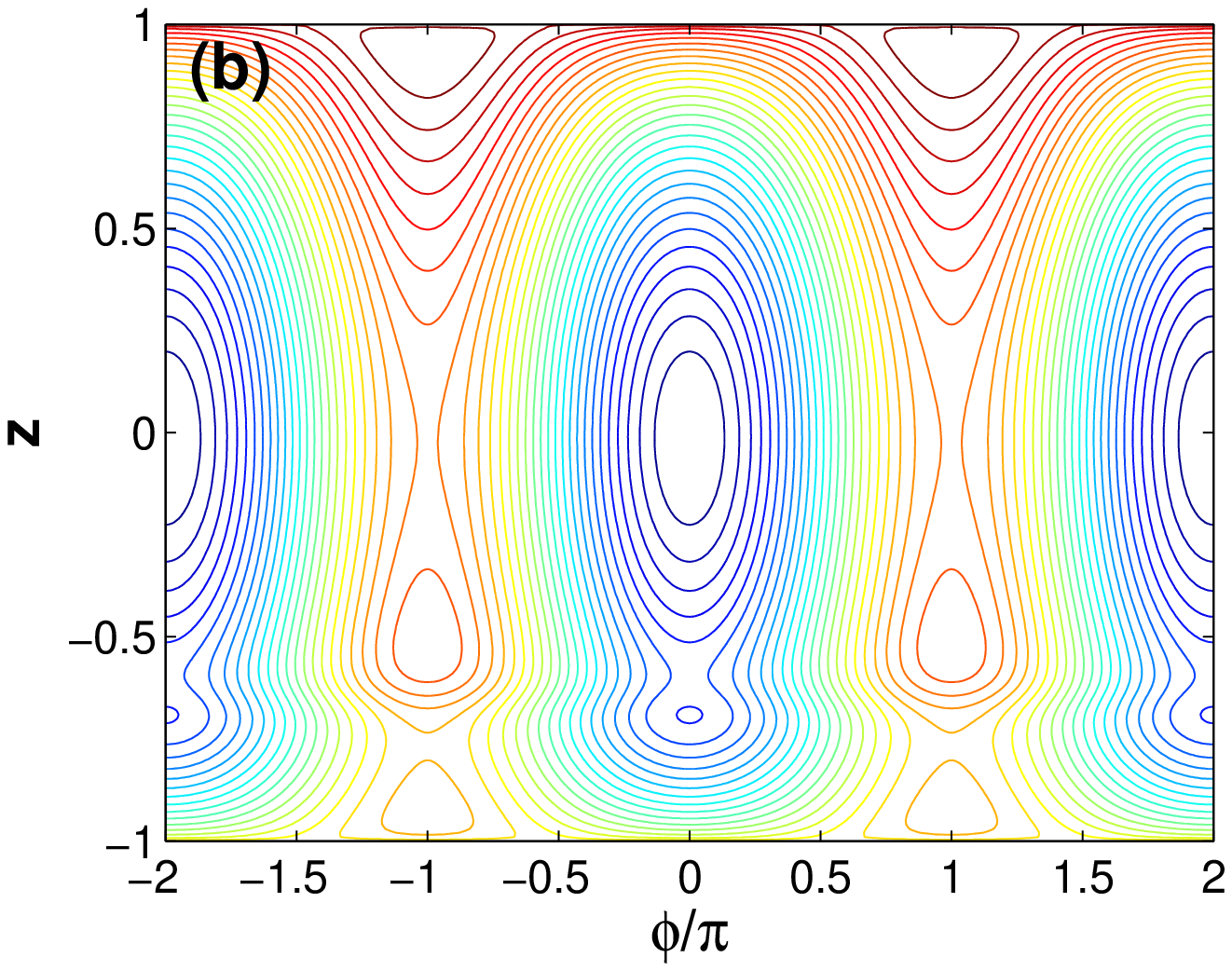}
\end{minipage}\\
\begin{minipage}[b]{0.45 \textwidth}
\includegraphics[width=\textwidth]{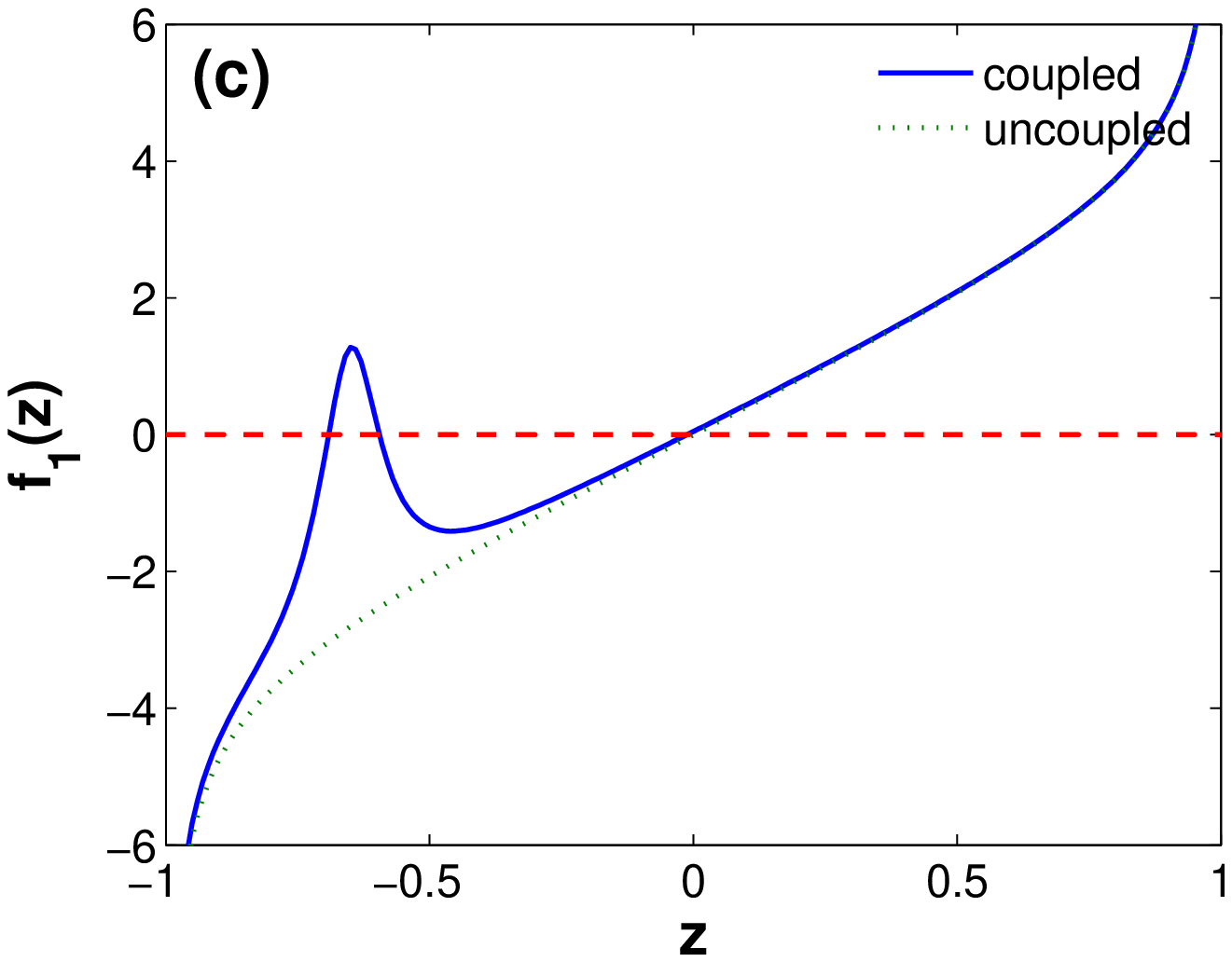}
\end{minipage}
\begin{minipage}[b]{0.45 \textwidth}
\includegraphics[width=\textwidth]{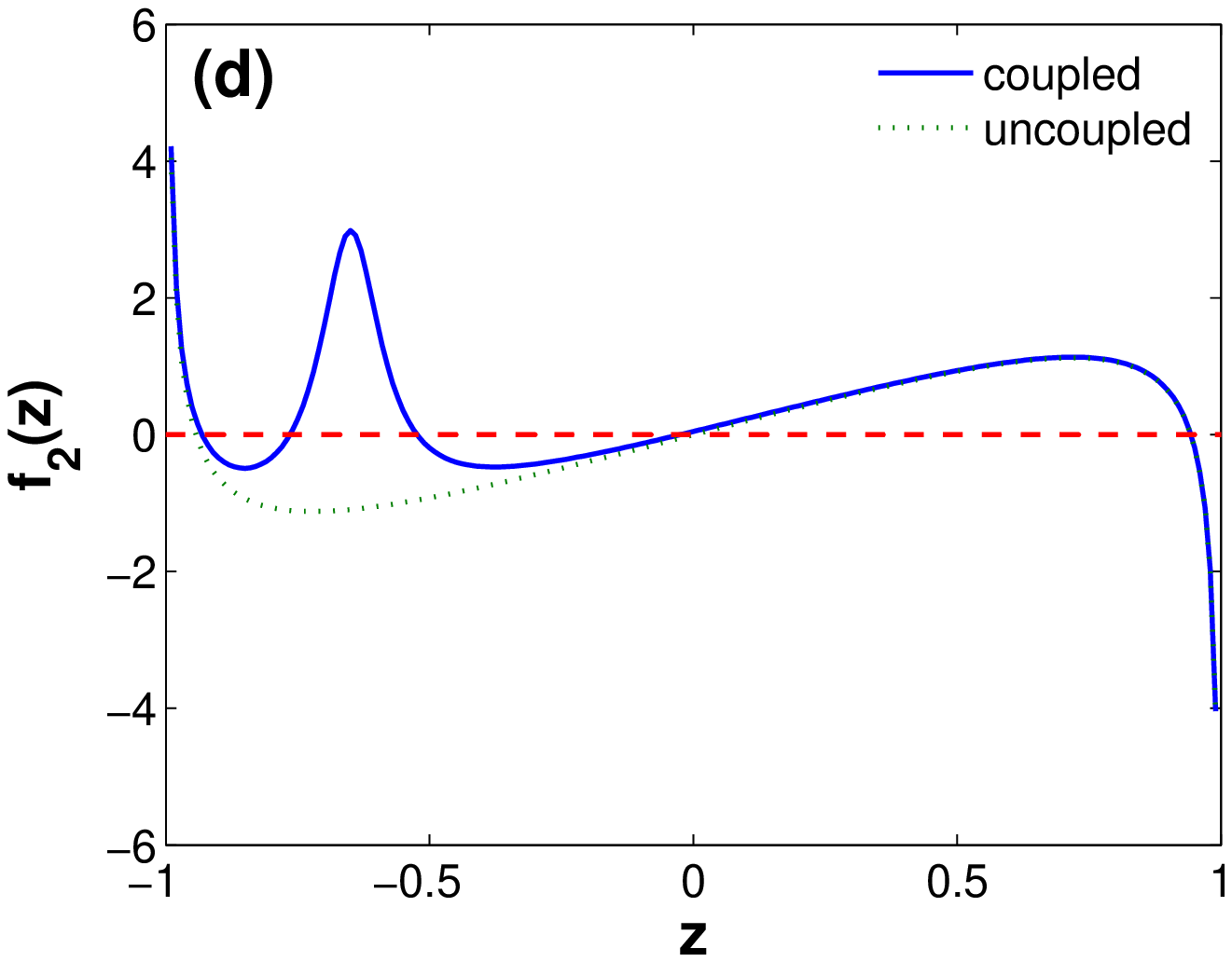}
\end{minipage}
\caption{\label{fig1}(Color online) Energy contours of a Bose
Josephson junction (a) uncoupled and (b) coupled to a single cavity
mode. (c) and (d): Gradient of the energy along the line $\phi=0$
and $\phi=\pi$, respectively. Zeros of $f_1(z)$ with positive
(negative) derivatives correspond to minima (saddle points) of
$H_c$, while zeros of $f_2(z)$ with positive (negative) derivatives
correspond to saddle points (maxima). The parameters are
$NV/(2\Omega)\equiv r=3$, $\tilde{A}=0.02$, $B=-0.65$, and $C=0.07$
\cite{feasibility}.}
\end{figure*}

For the coupled case, the roots of Eqs.~(\ref{z1}) and (\ref{z2})
have to be solved numerically. It is natural to expect that the last
term will not only shift the positions of the stationary points, but
may also alter the total number of them. Thus the phase portrait of
a cavity field coupled BJJ may be quantitatively or even
qualitatively different from that of the uncoupled case. A
particular example is given in Fig.~\ref{fig1}. As shown in
Figs.~\ref{fig1}(c)-(d), in the specific set of parameters, both the
functions $f_1(z)$ and $f_2(z)$ have two new roots when the BJJ
couples to the single cavity mode. The two new roots of $f_1(z)$
give rise to a new minimum and a new saddle point along the line
$\phi=0$, while those of $f_2(z)$ correspond to a new maximum and a
new saddle point along the line $\phi=\pi$, as clearly visible in
the contour map of $H_c$ in Fig.~\ref{fig1}(b). Comparing Fig.~1(b)
with Fig.~1(a), we see that the cavity mode coupled BJJ has more
complex and diverse behaviors than its uncoupled counterpart. To be
specific, the coupled BJJ has now three types of zero-phase modes
and three types of $\pi$-phase modes, while the uncoupled BJJ
possesses just one type of zero-phase mode and two types of
$\pi$-phase modes. We attribute the appearance of new stationary
points, and hence new motional modes of the BJJ, to the nonlinearity
of the cavity field induced tilt. To appreciate this point, let us
consider the tilt due to the zero-point energy difference of the two
traps or height difference in the gravitational field. That will
contribute a term linear in $z$ to the Hamiltonian $H_c$
\cite{smerzi, smerzi2}, and in turn a constant to the functions
$f_{1,2}$. As a constant just shifts the graph of a function up or
down as a whole, it is ready to convince oneself that no new roots
will arise.
\begin{figure}[t]
\centering
\includegraphics[bb=10 25 322 235, width=0.5\textwidth]{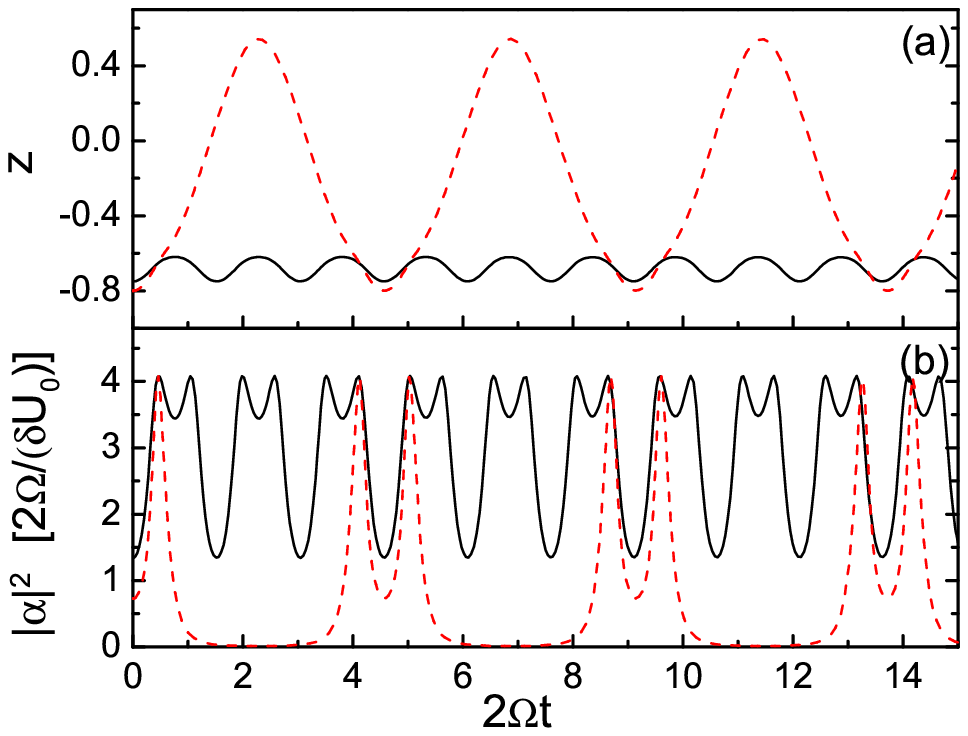}
\includegraphics[bb=10 25 322 225, width=0.5\textwidth]{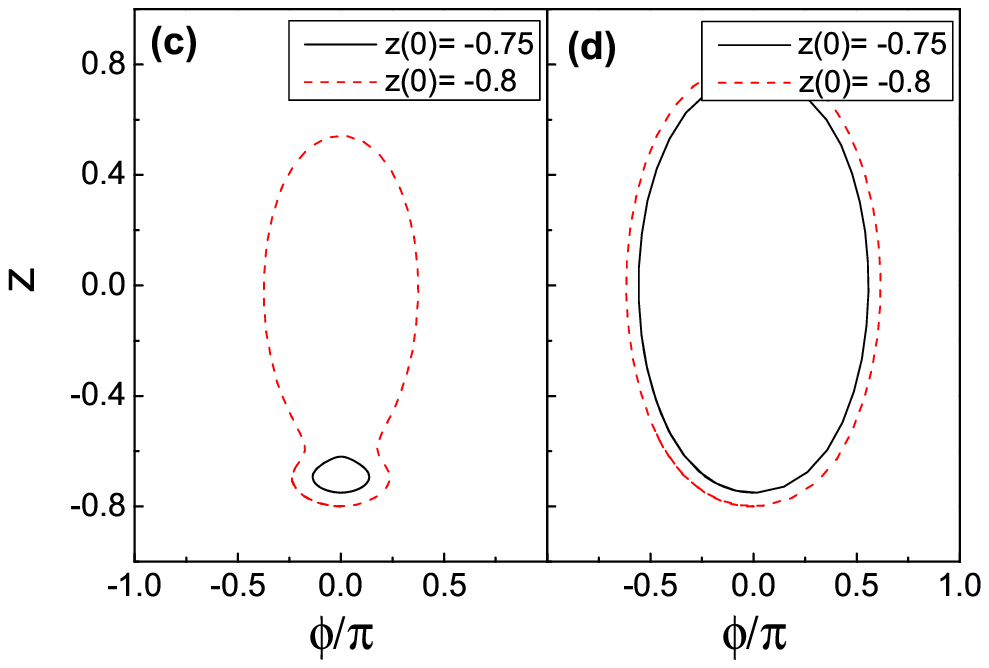}
\caption{\label{fig2}(Color online) Motion of the Bose Josephson
junction. (a) Atom population imbalance $z$ and (b) intra-cavity
photon number $|\alpha|^2$ [in units of $2\Omega/(\delta U_0)$]
versus the reduced time $2\Omega t$. The initial conditions are
$(\phi(0), z(0))=(0,-0.75)$ (black solid lines) and $(\phi(0),
z(0))=(0,-0.8)$ (red dashed lines). (c) and (d): trajectories of the
Bose Josephson junction (c) coupled or (d) uncoupled to the cavity
mode, with the two different initial conditions above. The same
parameters as in Fig.~\ref{fig1}.}
\end{figure}

We note that the cavity mode plays a dual role here. On one hand, it
plays with the condensates interactively and modifies their dynamics
effectively; on the other hand, it also carries with it the
information of the population of the atoms between the two traps as
it leaks out of the cavity. In Figs.~\ref{fig2}(a)-(b), we plot the
time evolution of the population imbalance and the number of
intra-cavity photons (which is proportional to the cavity output),
with the latter calculated from the former by using Eq.~(\ref{f}).
Initially, the phase $\phi(0)=0$, and $z(0)=-0.75$, or $-0.8$,
respectively. Despite of the minimal difference between the two
initial states, the subsequent dynamics is quite different. The
outputs of the cavity differ not only in their periods, but also in
their detailed temporal structures. The trajectories of the BJJ, as
can be read off from Fig.~\ref{fig1}(b), are shown in
Fig.~\ref{fig2}(c). The influence of the cavity field on the BJJ
dynamics can be seen by comparing Fig.~\ref{fig2}(c) with
Fig.~\ref{fig2}(d). It is worth noting that this influence may occur
at an extremely low intra-cavity photon number \cite{feasibility,
gupta}. If the parameters can be determined independently, we may
infer the population imbalance evolution from the outputs of the
cavity. This may serve as a different approach, which is
nondestructive, than the usual absorption image method, to track the
tunneling dynamics of two weakly linked Bose-Einstein condensates.
Of course, because the atom-field interaction involves only the atom
numbers [see Eq.~(\ref{Hint}) or Eqs.~(\ref{asolve}) and
(\ref{photonnumber})], no information of the relative phase of the
two condensates is contained in the cavity outputs. To fully
characterize the dynamics of a BJJ, techniques such as the
release-and-interfere \cite{schumm,albiez, shin} are still needed.

So far, we have assumed that the condensates remain intact during
their interaction with the cavity mode. One concern is that, due to
the temporal fluctuations of the intra-cavity photon number and
hence the fluctuations of the intra-cavity optical lattice, some
atoms may be diffracted into higher momentum modes \cite{meystre}
(this effect is sacrificed artificially here because of the two-mode
approximation we adopt). However, thanks to the fact that the mean
intra-cavity photon number here is very low (on the order of $0.01$,
see \cite{feasibility}), the strength of the fluctuations of the
optical lattice is small, so that this effect may be negligible. In
fact, in Esslinger \textsl{et al}.'s experiment \cite{esslinger},
where the mean intra-cavity photon number was maintained below
$0.04$, which corresponded to a maximal lattice depth below $0.1$
recoil energy \cite{private}, no signatures of diffraction were
observed. Of course, the temporal fluctuations of the cavity field
may heat the condensates and lead to atom loss, as has been observed
experimentally \cite{colombe,murch}. In addition, it may also cause
phase diffusion of the BJJ. These effects may damp the Josephson
oscillation, and deserve further study.

In conclusion, we have derived an effective Hamiltonian for a Bose
Josephson junction dispersively coupled to a single cavity mode,
under the mean-field approximation. The change of the dynamics of
the Bose Josephson junction is studied by the means of phase
portraits. We gave just one example as in Fig.~\ref{fig1} for
illustrative purposes. However, it by no means exhausts all the
possibilities. In fact, by engineering the so many free parameters
in the Hamiltonian, a large variety of qualitatively different cases
are accessible. Although in this work, as a starting point, we have
restricted to the time-independent case, it may be interesting to go
into the time-dependent case. By using an external feedback
depending on the outputs of the cavity \cite{rempe2}, \textsl{in
situ} observation and manipulation of the state of a Bose Josephson
junction may be achieved. Furthermore, generalization of the present
scenario to a Josephson junction array, may be worth consideration.
That will allow us to study the cavity-mediated long-range
interactions \cite{rempe3} between far separated condensates.

This work was supported by NSF of China under Grant Nos.~90406017,
60525417, and 10775176, NKBRSF of China under Grant
Nos.~2005CB724508, 2006CB921400, 2006CB921206, and 2006AA06Z104. J.
M. Z. would like to thank S. Gupta and D.~M. Stamper-Kurn for their
kind help on understanding of their paper.


\begin{thebibliography}{99}
\bibitem{feasibility}
We think this set of parameters are experimentally practical. By
taking $2\Omega \sim 2\pi\times10^2$ Hz, $\kappa \sim2\pi\times
10^6$ Hz, $ U_0 \sim 2\pi\times10^5$ Hz, $\delta\sim 0.1$, $r \sim
10$, $N \sim 10^3$, $A \sim 10^{-2}$, we have $\tilde{A} \sim 0.01$,
$C\sim 0.1$, $|\alpha|^2 \sim 0.01$. Note that $B$ is free by
adjusting the cavity pump detuning.
\end{thebibliography}
\end{document}